\documentclass[aps,preprint,showpacs]{revtex4-1}

\usepackage{amssymb}
\usepackage{mathrsfs}
\usepackage{lipsum}
\usepackage{graphicx}
\usepackage[caption=false]{subfig}
\usepackage{mathtools}
\usepackage{epstopdf}
\usepackage{xcolor}
\usepackage{color}
\definecolor{atrovirens}{RGB}{100,160,0}
\usepackage[colorlinks, linkcolor=red, anchorcolor=green, citecolor=blue, urlcolor=atrovirens]{hyperref}
\DeclareGraphicsExtensions{.pdf,.jpg,.png,.eps}
\usepackage[toc,page,title,titletoc,header]{appendix}

\begin{document}
	
\title{Angular-spectrum-based analysis on the self-healing effect of Laguerre-Gaussian beams after an obstacle}

\author{Jian-Dong Zhang}
\affiliation{School of Physics, Harbin Institute of Technology, Harbin 150001, China}
\author{Zi-Jing Zhang}
\email[]{zhangzijing@hit.edu.cn}
\affiliation{School of Physics, Harbin Institute of Technology, Harbin 150001, China}
\author{Jun-Yan Hu} 
\affiliation{School of Physics, Harbin Institute of Technology, Harbin 150001, China}
\author{Long-Zhu~Cen} 
\affiliation{School of Physics, Harbin Institute of Technology, Harbin 150001, China}
\author{Yi-Fei Sun} 
\affiliation{School of Physics, Harbin Institute of Technology, Harbin 150001, China}
\author{Chen-Fei Jin}
\email[]{jinchenfei@hit.edu.cn}
\affiliation{School of Physics, Harbin Institute of Technology, Harbin 150001, China}
\author{Yuan Zhao}
\email[]{zhaoyuan@hit.edu.cn}
\affiliation{School of Physics, Harbin Institute of Technology, Harbin 150001, China}

\date{\today}
	
\begin{abstract}
Self-healing, as an exotic effect, has showed many potential applications.
In this paper, we focus on the self-healing effect of Laguerre-Gaussian beams after an obstacle.
By taking advantage of angular spectrum theory, we study self-healing limit of the beam against on-axis obstacle.
The dependence of self-healing capability on the radius of obstacle is analyzed.
Additionally, we briefly discuss the self-healing limit of the beam in an off-axis scenario. 
Our results indicate that field amplitude of the beam will be healed well when the obstacle is approximately on-axis without oversized radius, perhaps providing advantages for optical communication, imaging, and remote sensing systems.
\end{abstract}

\maketitle

\section{Introduction}
Within the past two decades, the orbital angular momentum (OAM) of light \cite{PhysRevA.45.8185} has developed rapidly and opened up a new avenue to a myriad of fields, such as quantum information processing \cite{PhysRevLett.115.070502,Fickler2014Interface,PhysRevLett.110.263602,PhysRevLett.109.040401}, optical communications \cite{Xi2015Quantum,Wang2012Terabit,Willner2015Optical,Krenn13648}, and optical sensing \cite{PhysRevLett.80.3217,Lavery2013Detection,PhysRevLett.112.200401,Zhang:18}.
Related to this, the methods for generation and measurement of OAM beams have gained a lot of attention.
The conventional generation devices include cylindrical lenses \cite{BEIJERSBERGEN1993123}, spiral phase plates \cite{TURNBULL1996183}, Q wave plates \cite{PhysRevLett.96.163905}, and computational holograms \cite{Heckenberg1992Generation}.
The measurement methods are varied in comparison, besides reverse deploy of above generation devices, one can use interferometry \cite{PhysRevLett.88.257901,PhysRevLett.112.153601,PhysRevLett.92.013601}, diffraction approach \cite{Mourka:11,PhysRevX.3.041031}, and geometric transformation \cite{PhysRevLett.105.153601,Malik2013Direct,Zhou2016Experimental}.

Physically, OAM of the beams is associated with the Poynting vector which is not parallel to the direction of propagation.
Mathematically, the beams carrying OAM---Laguerre-Gaussian (LG) and Bessel-Gaussian (BG) beams---are characterized by a spiral phase term of the form $\exp\left( i\ell \theta \right) $, where $\ell$ and $\theta$ are azimuthal mode index and coordinate, respectively.
Different azimuthal modes are orthogonal and thus form an infinite-dimensional Hilbert space \cite{PhysRevA.96.023829,Bouchard:18}, which provides a route to improve information capacity in a single photon.
Furthermore, recent years have seen many studies on other exotic properties of OAM beams, diffraction-free behavior \cite{doi:10.1080/0010751042000275259,Belyi:10}, rotational Doppler effect \cite{Lavery537,Lavery:14,PhysRevApplied.10.044014}, gradient force \cite{PhysRevLett.88.053601,PhysRevLett.91.093602}, to name a few.

As one of these properties, self-healing effect of OAM beams \cite{Mendoza-Hernandez:15,LITVIN20091078,Rop_2012,Nape:18,Fu:17,Fahrbach2012Propagation,Mclaren2014Self,PhysRevA.98.053818} has been topical of late,  as well as a fruitful testbed for conceptual or experimental demonstrations, optical communication \cite{Nape:18}, imaging \cite{Fahrbach2012Propagation}, remote sensing \cite{Fu:17},
and so on \cite{Mclaren2014Self,PhysRevA.98.053818}.
Of the previous studies, most aim to BG beams; by contrast, the study on LG beams is few and far between.
As the paraxial approximate solutions of free-space Helmholtz equation in cylindrical coordinates, LG modes can be generated efficiently and have numerous applications.
In addition, they show many unique properties and advantages when compared with BG beams.
Therefore, the study on self-healing effect of LG beams is as important as that of BG beams. 
However, present studies have different opinions on who has better self-healing capability between LG and BG beams \cite{Mclaren2014Self,Mendoza-Hernandez:15}.
This indicates that the related studies leave some gaps which need to bridge.
To this end, here we address the self-healing effect of LG beams after an obstacle.
Unlike previous protocols based upon intensity observation, we find a new way, angular spectrum theory, of analyzing this scenario.

The remainder of this paper is organized as follows. 
In Sec. \ref{s2}, we introduce the fundamental principle of self-healing effect.
Section \ref{s3} gives the angular spectrum theory, and the self-healing capability of an LG beam after an on- or off-axis obstacle is studied.
Finally, we summarize our work with a brief conclusion in Sec. \ref{s4}.

\section{Fundamental principles}
\label{s2}
\begin{figure*}[htbp]
	\centering
	\includegraphics[width=0.6\textwidth]{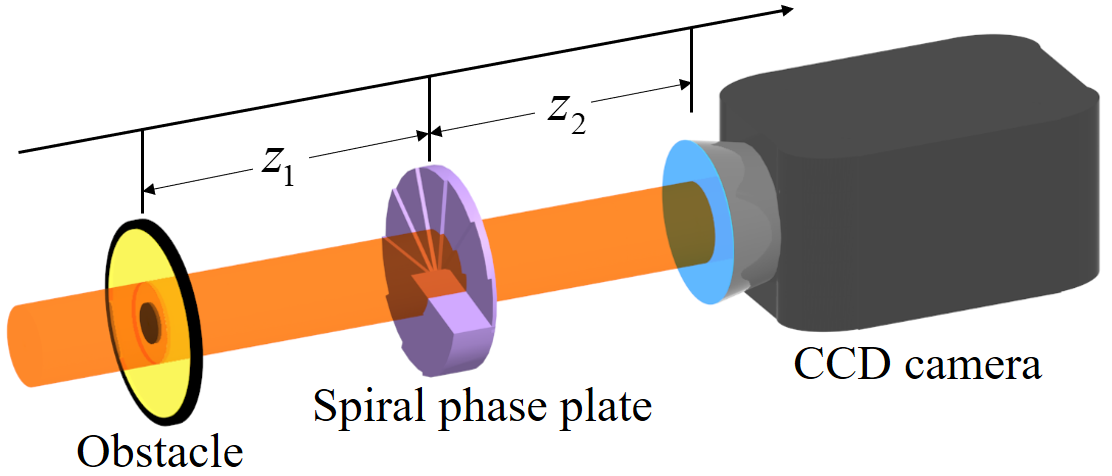}
	\caption{Schematic diagram of analysis on the self-healing effect of LG beams.}
	\label{system}
\end{figure*}

In what follows, we direct our attention to the schematic diagram of analysis on the self-healing effect, as shown in Fig. \ref{system}.
An OAM beam carrying $\ell$-fold spiral phase front is incident on an obstacle, an opaque disc inlaid in transparent glass.
After a propagation distance $z_1$, the beam is demodulated via a spiral phase plate (SPP) with reverse $\ell$-fold phase structure.
Finally, a CCD camera is placed at a distance of $z_2$ from the SPP to record intensity profile of the beam.
In principle, without the obstacle, the intensity profile presented by the CCD camera obeys Gaussian distribution for a large enough $z_2$.
However, once the obstacle appears on the direction of propagation, the regularity of field amplitude is broken.
Hand in hand with the amplitude destruction the beam starts repairing its field distribution.
The	primary task of this paper is to investigate the dependence of self-healing capability on the obstacle and propagation distance.

Consider an LG beam as illumination input, of which the normalized field amplitude can be expressed as:
\begin{eqnarray}
\nonumber E_p^\ell\left( {r,\theta ,z} \right) = &&\frac{{{D_{\ell p}}}}{\omega }{\left( {\frac{{\sqrt 2 r}}{\omega }} \right)^{\left| \ell \right|}}L_p^{\left| \ell \right|}\left( {\frac{{2{r^2}}}{{{\omega ^2}}}} \right)\exp \left( { - \frac{{{r^2}}}{{{\omega ^2}}}} \right)\exp \left[ {\frac{{ik{r^2}z}}{{2\left( {{z^2} + z_R^2} \right)}}} \right] \\ 
&&\times \exp \left[ { - i\left( {2p + \left| \ell \right| + 1} \right)\phi } \right]\exp \left( {i\ell\theta } \right), 
\label{e0}
\end{eqnarray}
where $r$, $\theta$, and $z$ refer to cylindrical coordinates;
${D_{\ell p}} = \sqrt {{{2p!} \mathord{\left/
			{\vphantom {{2p!} {\pi \left( {p + \left| \ell \right|} \right)!}}} \right.
			\kern-\nulldelimiterspace} {\pi \left( {p + \left| \ell \right|} \right)!}}} $
 is responsible for normalization;
$\phi  = \arctan \left( {{z \mathord{\left/
			{\vphantom {z {{z_R}}}} \right.
			\kern-\nulldelimiterspace} {{z_R}}}} \right)$ represents the Gouy phase;
${z_R} = {{k\omega _0^2} \mathord{\left/
		{\vphantom {{k\omega _0^2} 2}} \right.
		\kern-\nulldelimiterspace} 2}$ is known as the Rayleigh range with wave vector $k$;
${\omega _0}$ and $\omega  = {{{\omega _0}} \mathord{\left/ {\vphantom {{{\omega _0}} {\cos \phi }}} \right. \kern-\nulldelimiterspace} {\cos \phi }}$ are the beam waists at the beam focus and $z$ plane, respectively;	
$L_p^{\left| \ell \right|}\left( \cdot \right)$ stand for the associated Laguerre polynomials with $\ell$ and $p$ being the azimuthal and radial mode indices, respectively.

\begin{figure*}[htbp]
	\centering
	\includegraphics[width=0.8\textwidth]{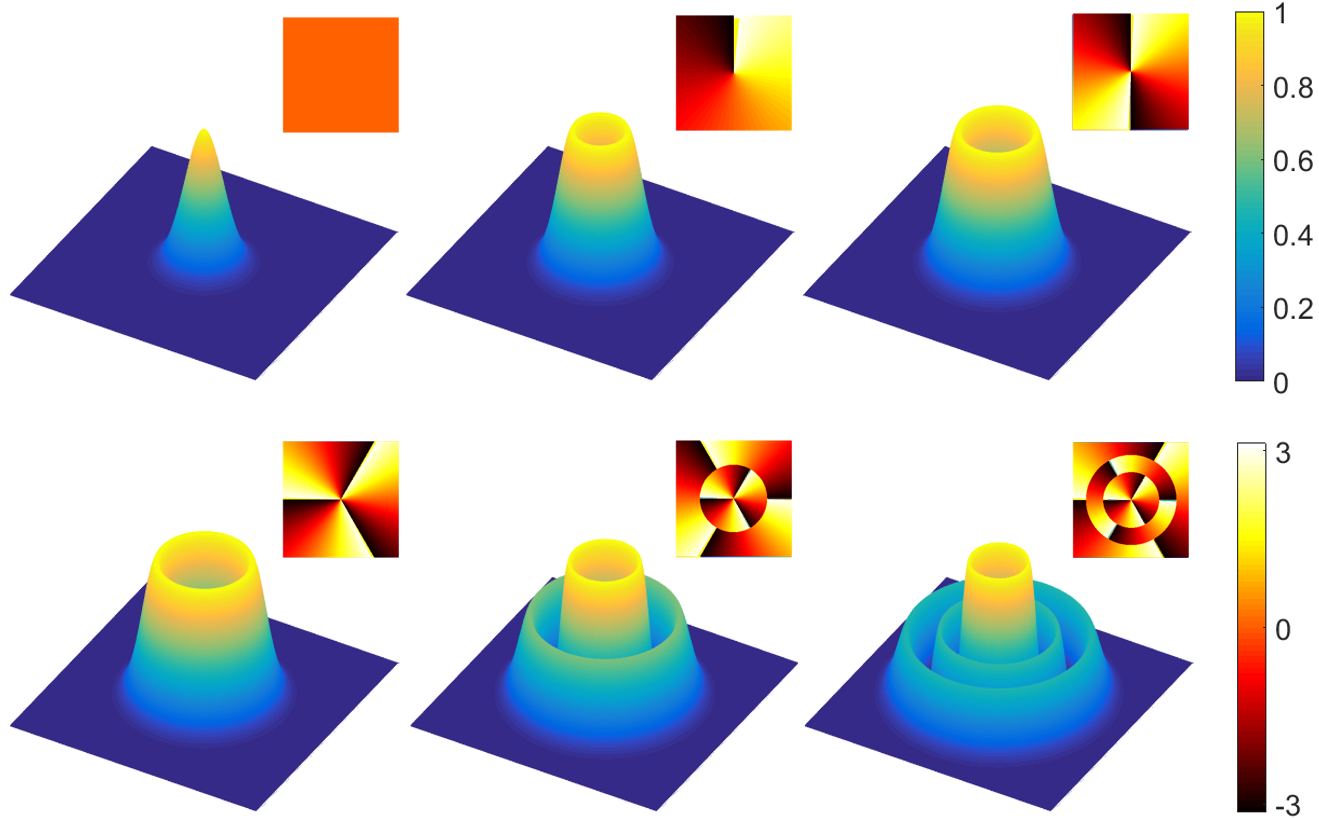}
	\caption{Intensity profiles of LG beams in the initial plane ($z=0$), where the upper right pattern in each inset is the corresponding phase front. 
	The top and bottom color bars on the right correspond to the intensity profiles and phase fronts, respectively.
	The mode indices chosen in the figure (from left to right): top row, $p = 0$ and $\ell = 0,1,2$; bottom row, $\ell = 3$ and $p = 0,1,2$.}
	\label{LG}
\end{figure*}

To such a field amplitude there corresponds a unique intensity profile with phase singularity, as illustrated in Fig. \ref{LG}.
Of most interest to us is the azimuthal mode index $\ell$, which describes the number of  phase jumps around the beam axis.
Throughout this paper, we focus on the LG beams with $p=0$ (doughnut beams \cite{Nicolas2013A}) as inputs.

Regarding the obstacle, it can be considered as an amplitude-only device; accordingly, the transfer function is given by
\begin{eqnarray}
{O\left( r \right)} = \left\{ {\begin{array}{*{20}{c}}
	{1{\kern 1pt} ,{\kern 1pt} {\kern 1pt} {\kern 1pt} {\kern 1pt} r \leqslant {r_0}{\kern 1pt} }  \\
	{0{\kern 1pt} ,{\kern 1pt} {\kern 1pt} {\kern 1pt} {\kern 1pt} r > {r_0}}  \\
	\end{array}} \right.
\end{eqnarray}
with ${r_0}$ being radius of the opaque disc.
It should be noted that, for an on-axis obstacle, an approximate expression without step functions can be used by the superposition of complex Gaussian functions \cite{doi:10.1121/1.396508}.
However, this approach is not adopted in this paper due to its intrinsic error along with error accumulation.

Upon leaving the obstacle, the output amplitude becomes ${E_0\left( {r,\theta,z } \right)} = {E\left( {r,\theta,z } \right)} {O\left( r \right)}$.
For simplicity, here we consider the LG beam in the initial plane, i.e., $z=0$.
When the beam arrives at the SPP through free-space propagation, its amplitude can be calculated from Collins diffraction integral equation \cite{Collins:70,Yang:13}
\begin{eqnarray}
\nonumber {E_1}\left( {{r_1},{\theta _1},{z_1}} \right) =&& \frac{i}{{\lambda {z_1}}}\exp \left( { - ik{z_1}} \right)\int_0^{2\pi } {\int_0^\infty  {E_0\left( {r,\theta } \right)} }  \\ 
&& \times \exp \left\{ { - \frac{{ik}}{{2{z_1}}}\left[ {{r^2} - 2{r_1}r\cos \left( {{\theta _1} - \theta } \right) + r_1^2} \right]} \right\}rdrd\theta, 
\label{e1} 
\end{eqnarray}
where $z_1$ is the propagation distance.

The SPP is a phase-only device, of which the transfer function can be described as $S\left( \theta_1 \right) =\exp \left( { - i\ell \theta_1} \right)$;
consequently, the amplitude will go as ${E'_1\left( {r_1,\theta_1,z_1 } \right)} = {E_1\left( {r_1,\theta_1,z_1 } \right)}\exp \left( { - i\ell \theta_1} \right)$ when the beam passes through the SPP. 
The analogy of Eq. (\ref{e1}) gives beam amplitude at the plane of CCD camera,
\begin{eqnarray}
\nonumber {E_2}\left( {{r_2},{\theta _2},{z_2}} \right) =&& \frac{i}{{\lambda {z_2}}}\exp \left( { - ik{z_2}} \right)\int_0^{2\pi } {\int_0^\infty  {E'_1\left( {r_1,\theta_1,z_1 } \right)} }  \\ 
&& \times \exp \left\{ { - \frac{{ik}}{{2{z_2}}}\left[ {{r_1^2} - 2{r_2}{r_1}\cos \left( {{\theta _2} - {\theta_1} } \right) + r_2^2} \right]} \right\}r_1dr_1d\theta_1.
\label{e2} 
\end{eqnarray}
Based on this amplitude expression, the self-healing effect can be analyzed.

\section{Results and discussions}
\label{s3}
\begin{figure*}[htbp]
	\centering
	\includegraphics[width=\textwidth]{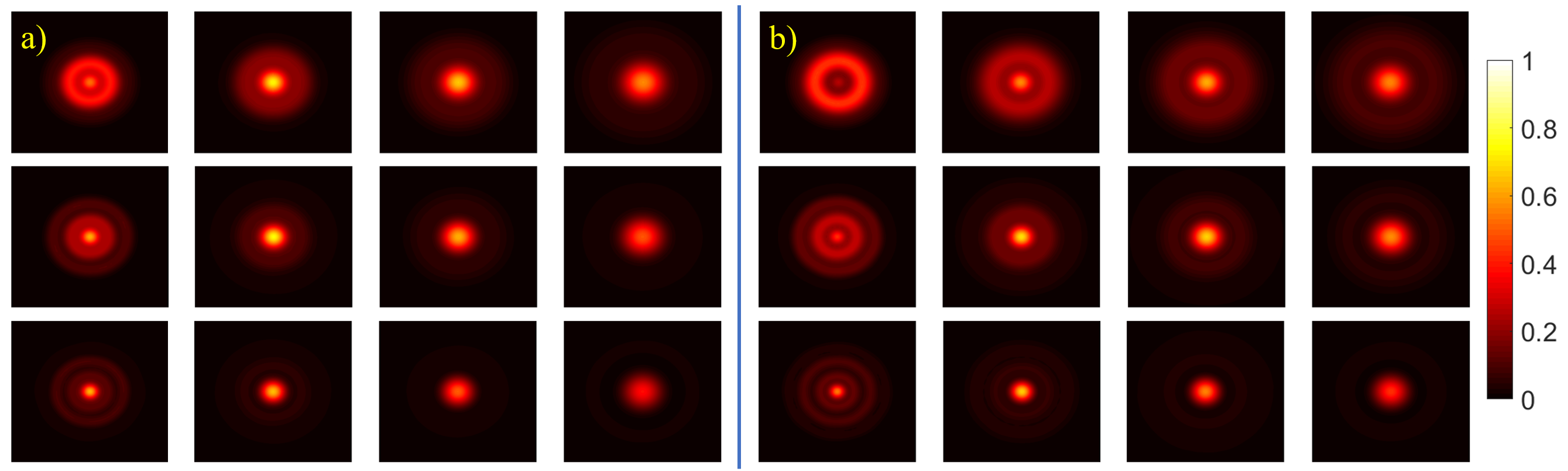}
	\caption{(a) Intensity profiles at the plane of CCD camera against different remainder intensity coefficients, $W$, and propagation distances, $z_2$, where $\ell=2$, $z_1=0.4$ m, and $\omega _0 = \sqrt {{\lambda  / \pi }}$ with wavelength $\lambda = 532$ nm. The dimensions of all subgraphs are 2 mm $\times$ 2 mm. From left to right: $z_2 =0.5$ m, 1 m, 1.5 m, 2 m; from top to bottom: $W=0.9$, 0.7, 0.5. (b) Same as group (a), but for $\ell=3$. The color bar on the right holds true for each subgraph.}
	\label{f3}
\end{figure*}

At the end of above section, the amplitude expression with overlap integral is given; however, it is by no means easy to free the amplitude expression from the integral.
Owing to the difficulty in providing a mathematically tractable expression, in this section, we utilize numerical approach to analyze the self-healing effect. 
In Fig. \ref{f3}, we show intensity profiles at the plane of CCD camera with different remainder intensity coefficients, $W$, and propagation distances, $z_2$.
The definition of remainder intensity coefficient is $W = {{\int_0^{2\pi } {\int_0^\infty  {{{\left| {{E_0}} \right|}^2}rdrd\theta } } } \mathord{\left/
		{\vphantom {{\int_0^{2\pi } {\int_0^\infty  {{{\left| {{E_0}} \right|}^2}rdrd\theta } } } {\int_0^{2\pi } {\int_0^\infty  {{{\left| E \right|}^2}rdrd\theta } } }}} \right.
		\kern-\nulldelimiterspace} {\int_0^{2\pi } {\int_0^\infty  {{{\left| E \right|}^2}rdrd\theta } } }}$ (see Appendix for derivation), where $E$ and $E_0$ are the beam amplitudes before and after the obstacle, respectively.

It can be seen that intensity profiles converge towards Gaussian mode with the increase of propagation distance.
By comparing with Figs. \ref{f3}(a) and \ref{f3}(b), one can also find that a high-order LG beam needs a larger propagation distance to approximately evolve into Gaussian mode.
For a small propagation distance, several rings appear in the outer region of the beam. 
These rings gradually blur as the beam continues to approach the plane of CCD camera.
Regarding the same propagation distance, the multi-ring phenomenon becomes more obvious when decreasing remainder intensity coefficient, i.e., the increase of radius of obstacle.
In addition, to accelerate the evolution of an OAM beam to Gaussian mode, one can add a lens after the SPP, or directly use a so-called vortex lens \cite{Swartzlander06,PhysRevApplied.7.034010} in place of the SPP.

In order to explain the origin of multi-ring phenomenon and explore the self-healing capability, we apply angular spectrum theory \cite{Torner:05,Chen2014Quantum} to analyze the self-healing effect of LG beams.
According to this theory, the amplitude of an arbitrary OAM beam can be decomposed into LG modes, for different LG modes form an infinite-dimensional orthogonal basis.
In this way, the amplitude after the obstacle can be written as:
\begin{eqnarray}
{E_0}\left( {r,\theta } \right) = \frac{1}{{\sqrt {2\pi } }}\sum\limits_\ell^{} {{a_\ell}} \left( r \right)\exp \left( {i\ell\theta } \right),
\end{eqnarray}
where the weight factor can be obtained by the discrete Fourier relationship \cite{Yao:06}
\begin{eqnarray}
{a_\ell}\left( r \right) = \frac{1}{{\sqrt {2\pi } }}\int_0^{2\pi } {{E_0}\left( {r,\theta } \right)} \exp \left( { - i\ell\theta } \right)d\theta. 
\end{eqnarray}

Further, we can determine the probability of finding a photon with an OAM of $\ell \hbar $ in the beam after the obstacle,
\begin{eqnarray}
{P_\ell} = \frac{{{C_\ell}}}{{\sum\nolimits_\ell {{C_\ell}} }}
\end{eqnarray}
with
${C_\ell} = \int_0^\infty  {{{\left| {{a_\ell}\left( r \right)} \right|}^2}} rdr$.
The self-healing capability refers to the degree to which the beam can maintain its original form.
As a result, we take advantage of the normalized probability $P_\ell$ to delineate self-healing capability.

\begin{figure*}[!bp]
	\centering
	\includegraphics[width=0.9\textwidth]{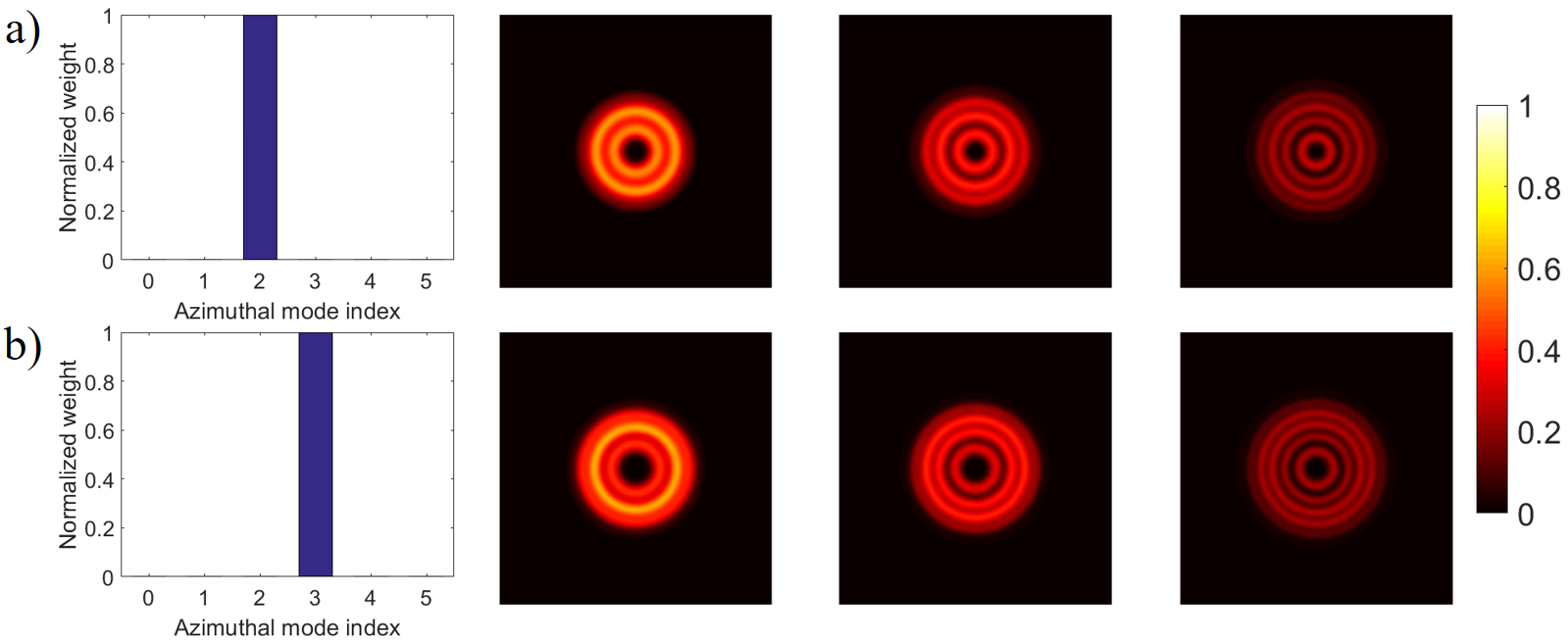}
	\caption{(a) Angular spectra and intensity profiles before the SPP against different remainder intensity coefficients $W$, where $\ell=2$, $z_1=0.4$ m, and $\omega _0 = \sqrt {{\lambda  / \pi }}$ with wavelength $\lambda = 532$ nm. The dimensions of all intensity subgraphs are 2 mm $\times$ 2 mm. From left to right: $W =0.9$, 0.7, 0.5. (b) Same as group (a), but for $\ell=3$. The color bar on the right holds true for each subgraph of intensity profiles.}
	\label{f4}
\end{figure*}

After angular spectrum calculation, we can find that the spectrum after an on-axis obstacle stays the same irrespective of the radius of obstacle, as depicted in Fig. \ref{f4}.
Meanwhile, a large radius of obstacle corresponds to obvious multi-ring phenomenon when compared with a small one.
The physical reason behind these is the spread in spectrum resulting from the obstacle.
On the one hand, the rotational symmetry of the obstacle prevents the changes in azimuthal indexes and further enforces angular spectrum conservation.
On the other hand, due to the limitation in the radial direction, different radial modes will interact and change \cite{PhysRevA.96.023829}.
Based on the above statement, the field amplitude after the obstacle can be described as a superposition of LG modes,
\begin{eqnarray}
{E_1}\left( {{r_1},{\theta _1},{z_1}} \right) = \sum\limits_{p = 0}^\infty  {{\alpha _p}E_p^\ell\left( {{r_1},{\theta _1},{z_1}} \right)},
\end{eqnarray}
all with the same azimuthal index but a range of radial indices \cite{Yao:11,DENNIS2009293}.

\begin{figure*}[htbp]
 	\centering
 	\includegraphics[width=0.8\textwidth]{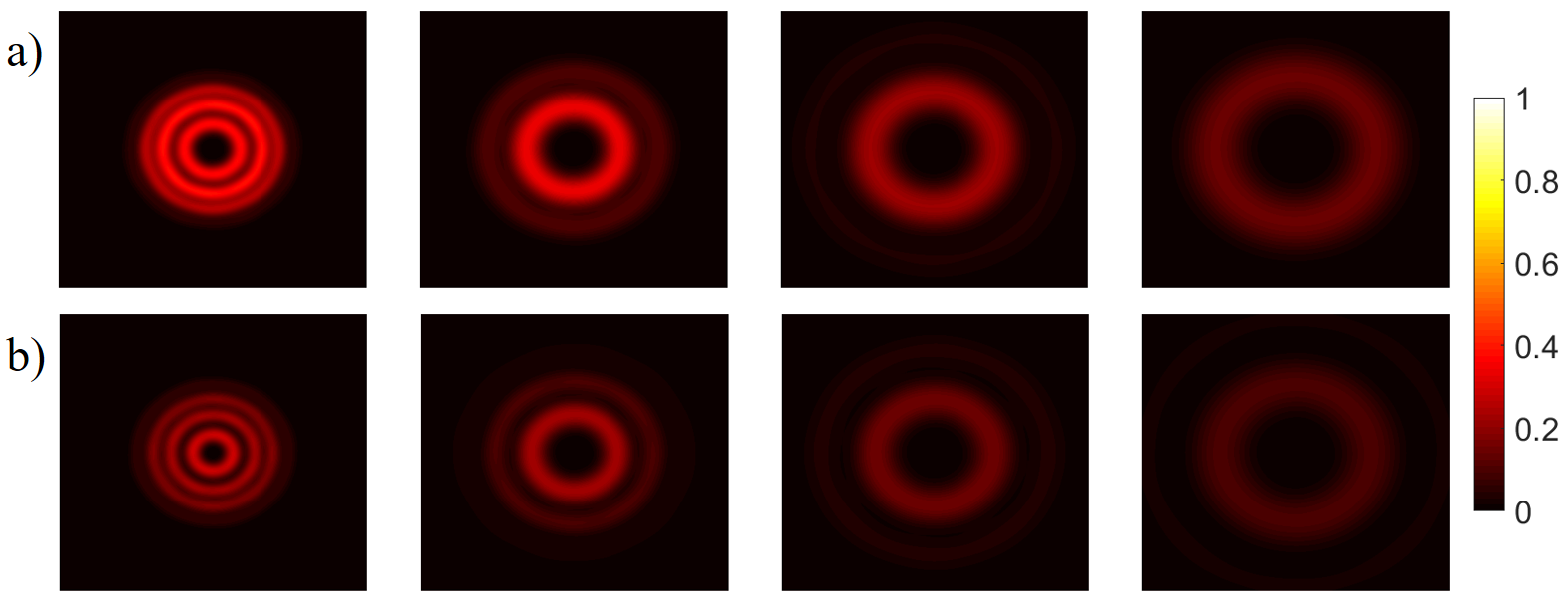}
 	\caption{(a) Intensity profiles after the obstacle against different propagation distances $z_1$, where $\ell=3$, $W=0.7$, and $\omega _0 = \sqrt {{\lambda  / \pi }}$ with wavelength $\lambda = 532$ nm. From left to right: $z_1 =0.5$ m, 1 m, 1.5 m, 2 m. The dimensions of all subgraphs are 2 mm $\times$ 2 mm. (b) Same as group (a), but for $W=0.5$. The color bar on the right holds true for each subgraph of intensity profiles.}
 	\label{f5}
 \end{figure*}

With different propagation distances and radii taken, Fig. \ref{f5} shows the intensity profiles after the obstacle. 
The results suggest that either decreasing radius or increasing propagation distance can weaken the multi-ring phenomenon.
This means that the field amplitude converge towards a nearly pure LG beam in the far field.
Overall, however great the radius may be, an LG beam can always heal its broken field amplitude as the angular spectrum remains the same.

Finally, we briefly discuss the self-healing capability in the presence of an off-axis obstacle.
At this point, the transfer function of the obstacle is recast as:
\begin{eqnarray}
{O' \left( r \right)}= \left\{ {\begin{array}{*{20}{c}}
	{1{\kern 1pt} ,{\kern 1pt} {\kern 1pt} {\kern 1pt} {\kern 1pt} \left| {r - \varepsilon } \right| \leqslant {r_0}{\kern 1pt} }  \\
	{0{\kern 1pt} ,{\kern 1pt} {\kern 1pt} {\kern 1pt} {\kern 1pt} \left| {r - \varepsilon } \right| > {r_0}}  \\
	\end{array}} \right.
\end{eqnarray}
with an off-axis displacement $\varepsilon$, and other calculations are the same.
Here the transfer function is labeled with a prime so as to distinguish it from that in on-axis scenario.

For an off-axis obstacle, it is invalid to simply define the remainder intensity coefficient, $W$.
Hence, given a beam waist at the beam focus $\omega_0$, we introduce two dimensionless parameters as follows: radial coefficient $R_1 = r_0 / \omega_0$, and displaced coefficients $R_2 = \varepsilon / \omega_0$.

\begin{figure*}[htbp]
	\centering
	\includegraphics[width=0.8\textwidth]{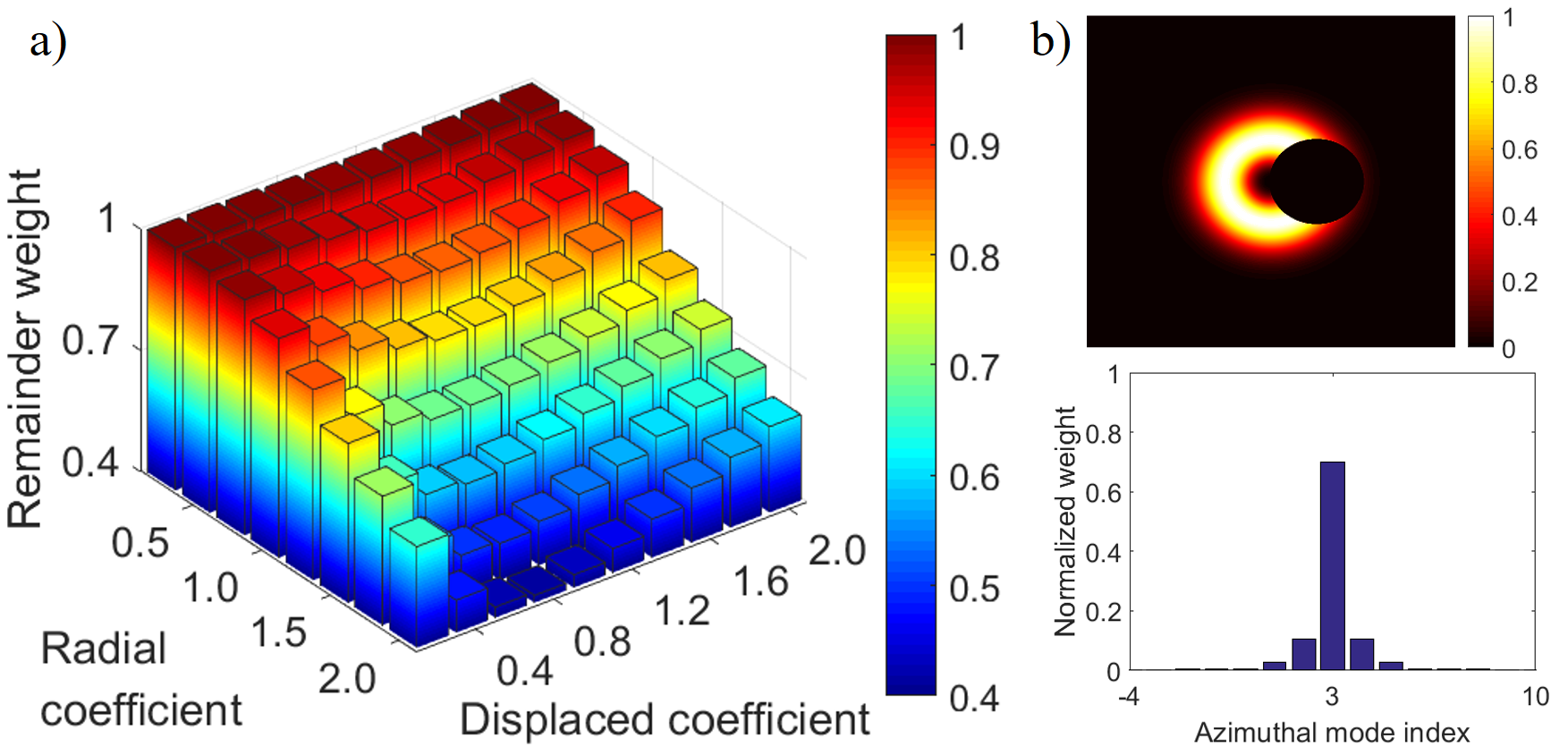}
	\caption{(a) Remainder weights $P_\ell$ against different radial coefficients, $R_1$, and displaced coefficients, $R_2$, where $\ell=3$ and $\omega _0 = \sqrt {{\lambda  / \pi }}$ with wavelength $\lambda = 532$ nm. (b) Intensity profile and angular spectrum in the case of $R_1=1.25$ and $R_2=1.2$.}
	\label{f6}
\end{figure*}

Figure \ref{f6} gives the dependence of remainder weight, $P_\ell$, on radial coefficient, $R_1$, and displaced coefficient, $R_2$.
One can find that the effect of a small radial coefficient on spread in spectrum is always negligible (${P_\ell} \approx 1$) regardless of displaced coefficient.
Meanwhile, for a given radial coefficient, remainder weight is not a monotonic function of displaced coefficient.
The most serious spread in spectrum will occur when a segment of the beam intensity is completely shielded, as shown in Fig. \ref{f6} (b).
This originates from the maximum violation of rotational symmetry induced by the off-axis obstacle.
In addition, when the obstacle is approximately on-axis ($R_2 \leqslant 0.2$) and its radius is not oversized ($r_0 \leqslant \omega _0$ or $R_1 \leqslant 1$), the remainder weight will be superior to 90\%.

\section{Conclusions}
\label{s4}
In conclusion, we have studied the self-healing effect of LG beams.
The self-healing limit of the beam is discussed through the use of angular spectrum analysis.
Regarding to an on-axis obstacle, due to rotational symmetry, there is no spread in spectrum  no matter what radius of the obstacle is.
However, the interaction of radial modes becomes strong with the increase of radius; accordingly, non-zero radial mode cannot directly evolve into Gaussian mode.
In addition, with a radius of the obstacle given, the interaction of radial modes can be weakened by increasing the azimuthal index.
As to an off-axis obstacle, the weakest self-healing capability appear in the situation that a segment of the beam intensity is completely shielded.
For an approximately on-axis obstacle with a not oversized radius, $R_2 \leqslant 0.2$ and $R_1 \leqslant 1$, an initial angular spectrum of the beam in excess of 90\% can be well maintained.
These results may be useful for optical communication, imaging, and remote sensing systems.

\section*{Acknowledgment}
This work was supported by the National Natural Science Foundation of China (Grant No. 61701139).

\section*{Appendix}
\label{APP}
In this section, we provide the calculation approach to remainder intensity coefficient after an on-axis obstacle.
The result also applies to the scenario that an LG beam passes through an iris with a circular opening of radius $r_0$. 

First, one can verify that the field amplitude in Eq. (\ref{e0}) is normalized using the existing integral formula,
\begin{eqnarray}
\int_0^\infty  {{x^{2n + 1}}\exp \left( { - b{x^2}} \right)} dx = \frac{{n!}}{{2{b^{n + 1}}}}
\end{eqnarray}
That is, without any obstacles, the total intensity throughout space is 1. 
In addition, the specific expression of the associated Laguerre polynomial is found to be
\begin{eqnarray}
L_p^\ell\left( x \right) = \sum\limits_{k = 0}^p {{{\left( { - 1} \right)}^k}\frac{{\left( {\ell + p} \right)!}}{{k!\left( {p - k} \right)!\left( {\ell + k} \right)!}}} {x^k}.
\end{eqnarray}
It is obvious that the value of $k$ takes on 0 under the constraint of $p=0$; as a result, we have $L_p^\ell\left( x \right) =1$ for each value of variable $x$. 
Further, the corresponding intensity distribution can be simplified as: 
\begin{eqnarray}
{I_\ell}\left( {r,\theta ,z} \right) = \left| E_p^\ell\left( {r,\theta ,z} \right)  \right|^2= \frac{2}{{\pi \left| \ell \right|!}}\frac{1}{{{\omega ^2}}}{\left( {\frac{{\sqrt 2 r}}{\omega }} \right)^{2\left| \ell \right|}}\exp \left( { - \frac{{2{r^2}}}{{{\omega ^2}}}} \right).
\end{eqnarray}

Based on this expression, the remainder intensity of the beam after the obstacle can be calculated from the following integral
\begin{eqnarray}
W = 1-2\pi A\int_{0}^{r_0} {{r^{2\left| \ell \right| + 1}}\exp \left( { - a{r^2}} \right)} dr
\label{e5}
\end{eqnarray}
with $A={{{2^{\left| \ell \right|{\rm{ + }}1}}} \mathord{\left/
		{\vphantom {{{2^{\left| l \right|{\rm{ + }}1}}} {\left( {\pi {\omega ^{2 + 2\left| l \right|}}\left| \ell \right|!} \right)}}} \right.
		\kern-\nulldelimiterspace} {\left( {\pi {\omega ^{2 + 2\left| \ell \right|}}\left| \ell \right|!} \right)}}$ and $a =  {2 \mathord{\left/
		{\vphantom {2 {{\omega ^2}}}} \right.
		\kern-\nulldelimiterspace} {{\omega ^2}}}$.
This can also be regarded as remainder intensity coefficient, for the input beam ahead of the obstacle is unit intensity. 
 
Using the following integral formula,
\begin{eqnarray}
\int {{x^n}\exp \left( { - b{x^2}} \right)dx =  - \frac{1}{2}\frac{1}{{{b^{{{\left( {n + 1} \right)} \mathord{\left/
							{\vphantom {{\left( {n + 1} \right)} 2}} \right.
							\kern-\nulldelimiterspace} 2}}}}}} \Gamma \left( {\frac{{n + 1}}{2},b{x^2}} \right),
\end{eqnarray}
we get 
\begin{eqnarray}
\nonumber W &&=  1 + \frac{1}{2}\frac{1}{{{a^{{{\left( {n + 1} \right)} \mathord{\left/
						{\vphantom {{\left( {n + 1} \right)} 2}} \right.
						\kern-\nulldelimiterspace} 2}}}}}\left[ {\Gamma \left( {\frac{{n + 1}}{2},a{{r_0}^2}} \right) - \Gamma \left( {\frac{{n + 1}}{2},0} \right)} \right] \\ 
\nonumber&&=  1 + \frac{1}{2}\frac{1}{{{a^{{{\left( {n + 1} \right)} \mathord{\left/
						{\vphantom {{\left( {n + 1} \right)} 2}} \right.
						\kern-\nulldelimiterspace} 2}}}}}\left[ {\Gamma \left( {\frac{{n + 1}}{2}} \right) - \int_0^{a{{r_0}^2}} {{t^{\frac{{n + 1}}{2} - 1}}{e^{ - t}}dt}  - \Gamma \left( {\frac{{n + 1}}{2}} \right)} \right] \\ 
&&= 1-\frac{1}{2}\frac{1}{{{a^{\left| \ell \right| + 1}}}}\int_0^{a{r_0^2}} {{t^{\left| \ell \right|}}{e^{ - t}}dt}.  
\label{e4}
\end{eqnarray}
Here we take $n = 2\left| \ell \right|{\rm{ + }}1$ and use the incomplete Gamma function and Gamma function
\begin{eqnarray}
\Gamma \left( {u,h} \right) = \int_h^\infty  {{t^{u - 1}}{e^{ - t}}dt = \Gamma \left( u \right)}  - \int_0^h {{t^{u - 1}}{e^{ - t}}dt},  \\ 
\Gamma \left( u \right) = \int_0^\infty  {{t^{u - 1}}{e^{ - t}}dt = } 2\int_0^\infty  {{t^{2u - 1}}\exp \left( { - {t^2}} \right)dt}.  
\end{eqnarray}
At this point, Eq. (\ref{e4}) can be rewritten as 
\begin{eqnarray}
W= 1 - \frac{1}{2}\frac{1}{{{a^{\left| \ell \right| + 1}}}}\left[ { - {e^{ - f}}\sum\limits_{n = 0}^\ell {\frac{{\ell!{f^{\ell - n}}}}{{\left( {\ell - n} \right)!}}} } \right]_{f = 0}^{f = a{r_0^2}}
\end{eqnarray}
with the integral formula
\begin{eqnarray}
\int {{x^n}{e^{bx}}dx}  = {e^{bx}}\sum\limits_{k = 0}^n {{{\left( { - 1} \right)}^k}\frac{{n!{x^{n - k}}}}{{\left( {n - k} \right)!{b^{k + 1}}}}}.
\end{eqnarray}
After some algebra, we obtain the expression of remainder intensity coefficient
\begin{eqnarray}
W = {\exp\left( { - a{r_0^2}}\right) }\sum\limits_{m = 0}^\ell {\frac{{{{\left( {a{r_0^2}} \right)}^{\ell - m}}}}{{\left( {\ell - m} \right)!}}}.
\end{eqnarray}
This means that the integral in Eq. (\ref{e5}) is converted into an finite series, which is easy to calculate.

\begin{figure*}[htbp]
	\centering
	\includegraphics[width=0.45\textwidth]{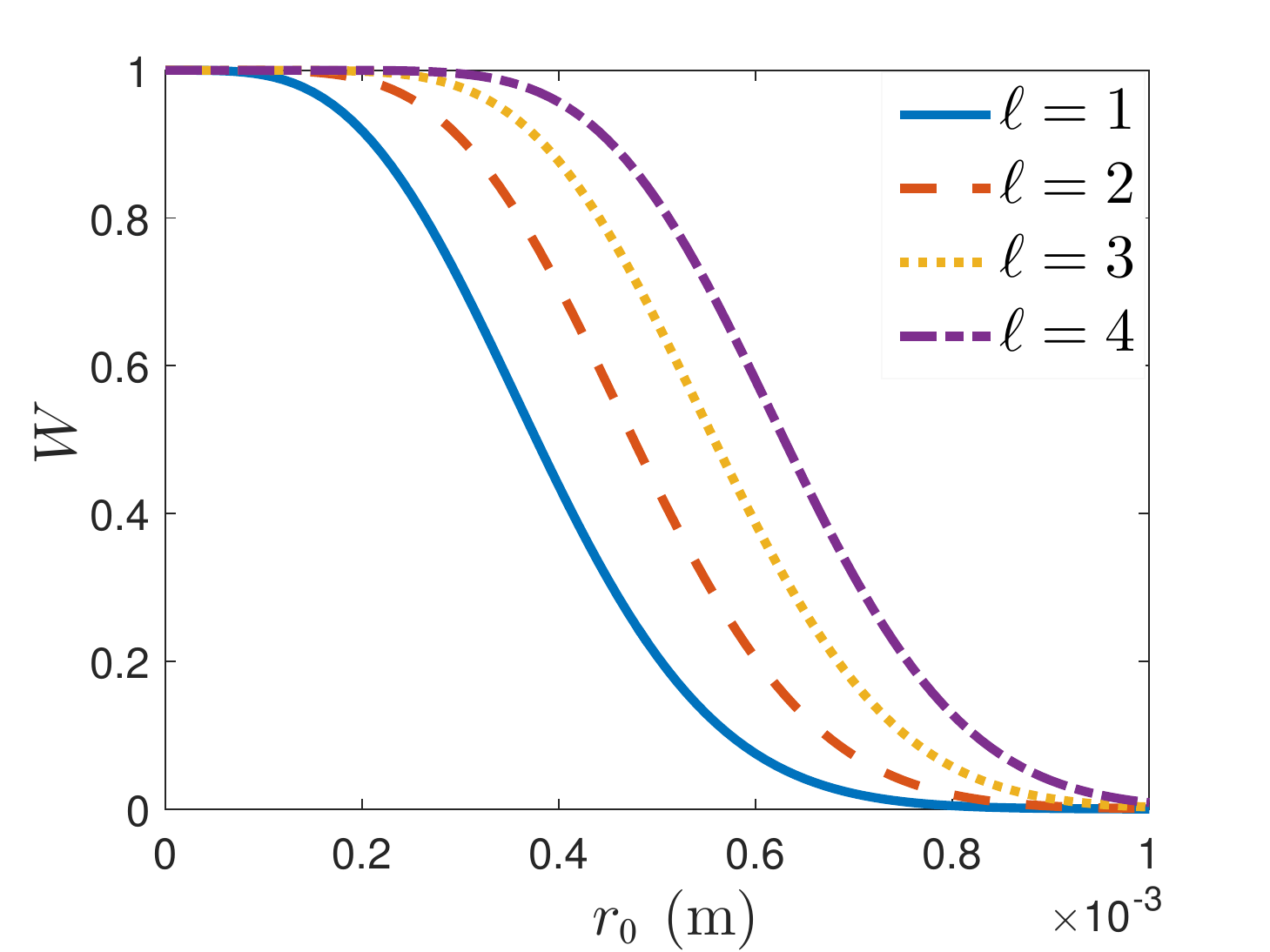}
	\caption{The remainder intensity coefficient versus the radius $r_0$ of the obstacle, where $a = 2/\sqrt {{\lambda  / \pi }}$ with wavelength $\lambda = 532$ nm.}
	\label{fig_end}
\end{figure*}

In Fig. \ref{fig_end} we show the remainder intensity coefficient $W$ as a function of the radius $r_0$ of the obstacle with different mode indices.
The result indicates that, for a given radius, a high-order LG beam corresponds to more  intensity since it has a large dark area compared with a low-order one. 

%

\end{document}